\documentclass[prl,superscriptaddress,reprint,showpacs,amsmath,amssymb,nofootinbib,nobalancelastpage]{revtex4-1}
\usepackage{amsmath,amssymb,graphicx,verbatim}
\usepackage{bbold,color} 
\usepackage[colorlinks=true,urlcolor=blue,linkcolor=blue]{hyperref}

\font\tenbsl=cmbxsl10

\def\({\left(}
\def\){\right)}

\def\a2d{a^{\dagger 2}}
\def\b2d{b^{\dagger 2}}

\def\fig#1{Fig.\ref{fig:#1}}

\def\sec#1{Section~\ref{sec:#1}}

\def\og{\leavevmode\raise.3ex\hbox{$\scriptscriptstyle\langle\!\langle$}}
\def\fg{\leavevmode\raise.3ex\hbox{$\scriptscriptstyle\,\rangle\!\rangle$}}

\def\refer#1#2{[{\tenbsl{#1}
\setbox100=\hbox{#2}\ifdim\wd100>10pt\kern .3em\box100$\,$\fi}]}

\def\lejour{le\ {\the\day}\
\ifcase\month\or janvier\or f\'evrier\or mars\or avril\or mai\or juin\or
juillet\or ao\^ut\or septembre\or octobre\or novembre\or d\'ecembre\fi\
{%\oldstyle
\the\year}}

\def\boxit#1#2{\setbox1=\hbox{\kern#1{#2}\kern#1}%
\dimen1=\ht1 \advance\dimen1 by #1 \dimen2=\dp1 \advance\dimen2 by #1
\setbox1=\hbox{\vrule height\dimen1 depth\dimen2\box1\vrule}%
\setbox1=\vbox{\hrule\box1\hrule}%
\advance\dimen1 by .4pt \ht1=\dimen1
\advance\dimen2 by .4pt \dp1=\dimen2 \box1\relax}

\usepackage{amsmath}
\usepackage[percent]{overpic}
\def\fig#1{Fig.\ref{fig:#1}}

\def\sec#1{Section~\ref{sec:#1}}

\begin{document}

\title{Photon-number-resolving segmented detectors based on single-photon avalanche-photodiodes}
\author{Rajveer Nehra}
\email{rn2hs@virginia.edu}
\author{Chun-Hung Chang}
\email{cc4us@virginia.edu}
\affiliation{Department of Physics, University of Virginia, 382 McCormick Rd,\\ Charlottesville, VA 22904-4714, USA}
\author{Qianhuan Yu}
\email{qy5pg@virginia.edu}
\affiliation{Department of Electrical and Computer Engineering, University of Virginia, 351 McCormick Rd, Charlottesville, VA 22903, USA}
\author{Andreas Beling}
\email{ab3pj@virginia.edu}
\affiliation{Department of Electrical and Computer Engineering, University of Virginia, 351 McCormick Rd, Charlottesville, VA 22903, USA}
\author{Olivier Pfister}
\email{opfister@virginia.edu}
\affiliation{Department of Physics, University of Virginia, 382 McCormick Rd,\\ Charlottesville, VA 22904-4714, USA}



\begin{abstract}
We investigate the feasibility and performance of photon-number-resolved photodetection employing single-photon avalanche photodiodes (SPADs) with low dark counts. While the main idea,  to split  $n$ photons into $m$ detection modes with a vanishing probability of more than one photon per mode, is not new, we investigate here a important variant of this situation where SPADs are side-coupled to the same waveguide rather than terminally coupled to a propagation tree. This prevents the nonideal SPAD quantum efficiency from contributing to photon loss. We propose a concrete SPAD segmented waveguide detector based on a vertical directional coupler design, and characterize its performance  by evaluating the purities of Positive-Operator-Valued Measures (POVMs) in terms of number of SPADs, photon loss, dark counts, and electrical cross-talk.
\end{abstract}
\maketitle
%%%%%%%%%%%%%%%%%%%%%%%%%%  body  %%%%%%%%%%%%%%%%%%%%%%%%%%
\section{Introduction}
 Quantum measurements are essential to quantum information science and technology. Photon-Number-Resolving (PNR) detection, in particular, fully exploiting the corpuscular nature of  classically undulatory light, is key in quantum metrology and sensing \cite{Quantum_metrology} and quantum technologies \cite{Migdall_book}.
  
A PNR detector produces a signal proportional to the number of incident photons. Photon-number-resolving detectors  have been realized with superconducting transition-edge sensors (TES) \cite{TES_2, Lita:08}, silicon photomultipliers \cite{Ramilli:10}, superconducting nanowires \cite{Divochiy:2008aa, Superconduncting_3, Superconducting_multiplexing_nanowire,Nicolich2019}, linear mode avalanche photodiodes (SPADs), and quantum-dot field-effect transistors \cite{doi:10.1063/1.2735281,doi:10.1063/1.126745}. Moreover, methods based on spatial- and time-multiplexing of non-PNR detectors have been proposed for PNR measurements using SPADs \cite{Kok2001,Achilles2003,Fitch03PRA, Piacentini:15, Superconducting_multiplexing_nanowire}. Such proposals have been thoroughly modeled mathematically~\cite{Sperling2012,Sperling2012subbinomial,Kruse2017,Miatto2018}.

\begin{figure}[h!]
\begin{center}
\begin{overpic}[width=0.78\columnwidth]{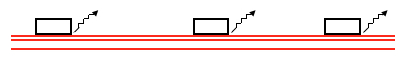}
 \put(82, 7.25){$\alpha_3$}
  \put(50, 7.2){$\alpha_2$}
 \put(12, 7.25){$\alpha_1$}
  \put(-3,2.8){$|n\rangle$}
  \put(19,12.5){\scriptsize loss channel}%$\sqrt{1-\alpha_1^2}$}
%   \put(57,12.5){\footnotesize$\sqrt{1-\alpha_2^2}$}
%    \put(89,12.5){\footnotesize$\sqrt{1-\alpha_3^2}$}
   \put(98,2.1){\scriptsize $\leftarrow$ waveguide} 
     \put(98,4.3){\scriptsize $\leftarrow$ cladding}  
 \end{overpic}
 \caption{Principle sketch of a segmented detector. Guided photons are detected alongside propagation by SPADs which frustrate total internal reflection. The quantum efficiency (QE) of SPAD \#$j$ is $\alpha_{j}^{2}$. The design goal is to eschew detection losses, which are distinct from the nonunity of $\alpha_{j}^{2}$, and keep all undetected photons in the waveguide for further detection. %Nonunity SPAD QE won't contribute to losses in this design.
 }{\label{waveguide}}
 \end{center}
 \end{figure}

In this paper we investigate the possibility of PNR detection using a segmented detector, constituted by waveguide-coupled, low-dark-current SPADs, as per Fig.~\ref{waveguide}.  
This linear array is essentially a long detector divided into $m$ detector segments, each with an individual read-out. The gist of this design is that photons that are not absorbed in the first SPAD must not be lost and be coupled back into the waveguide to be absorbed later. The crucial advantage of this configuration is that nonideal quantum efficiency of the SPADs doesn't amount to photon loss, unlike terminally coupled PNR detectors in which temporally or spatially split photons impinge on SPADs on the end of their path ~\cite{Fitch03PRA,Achilles2003,Piacentini:15}. Moreover, the SPAD coupling should follow a gradient down the waveguide so as to ensure no more than one photon is detected at a time (since SPADs are not PNR) while still ensuring efficient detection. The design goal is therefore to whittle down an initial $n$ photons, one by one.   We envision that such a segmented photodetector will become feasible in large-scale integrated photonic  platforms using either monolithic or heterogeneous integration of SPADs on low-loss waveguides, as has already been hinted at by the integration on waveguides of PIN photodiodes~\cite{Wang2017,Yu:19} and of transition edge sensors~\cite{Hopker2017,Hopker2019}.

\begin{figure}[h!]
\centerline{\begin{overpic}[width = 0.4\textwidth]{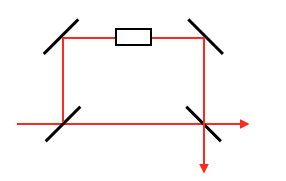}
 \put (90,20) {$tt' + rr'\sqrt{1-\alpha^2}  $}
 \put(60,-3){$ =
 tr' - rt'\sqrt{1-\alpha^2} $}
 \put(-4,18){$|n\rangle$}
 \put(22,13){$(r,t)$}
 \put(57,13){$(r',t')$}
 \put(45, 49){$\alpha$}
\end{overpic}}
\caption{Model for detection alongside propagation. If this is the $\rm j^{th}$ SPAD, then $t_{j}=tt' + rr'\sqrt{1-\alpha^2}$, where $t_{j}=T_{j}^{1/2}$ in \fig{seg}.}{\label{fig:detail}}
 \end{figure}

The essential physics of the SPAD coupling can be captured by a simplified model, pictured in Fig.~\ref{fig:detail}, which assumes that the SPAD length is exactly equal to the period of the mode beat  between the main waveguide and the SPAD. In \sec{wg}, we give concrete and detailed waveguide modeling results for this configuration, which has already been experimentally realized for PIN photodiodes~\cite{Yu:18}. The simplified model will be enough, without loss of generality, for the quantum analysis of the PNR behavior in \sec q. We take the SPAD 
quantum efficiency to be $\alpha^{2}$, accounting for both coupling efficiency and intrinsic absorption, such that its field transmissivity is $1-\alpha^{2}$. Note that it is desirable for $\alpha$ not be too large, so that the probability for any SPAD to see more than one photon during the same detection window can be vanishing, since SPADs are not PNR detectors; this translates into the condition $\alpha^2\ll 1/n$, for $n$ incident photons. In some cases, the mere click statistics from  a click/no-click detector system suffice to certify the non-classicality of a state~\cite{Sperling2012, Lee2016}. In such cases, the proposed design is particularly beneficial as it increases the overall detection efficiency by recycling the photons which are not absorbed at the first time but are detected as they propagate in the waveguide.
We require that the bottom output of the exit beamsplitter in Fig.~\ref{fig:detail}, which is effectively the radiative loss channel of the waveguide, be nulled by destructive interference. The condition can be achieved in the presence of SPAD absorption by choosing parameters $(r,t,r',t')$ of the beamsplitters, and absorption coefficient $\alpha$, such that 
\begin{equation}\label{l}
tr' - rt'\sqrt{1-\alpha^2} = 0.
\end{equation}
If this is the case, then the detection process truly takes place alongside propagation and finite quantum efficiency --- necessary here to attain PNR performance with SPADs, by detecting no more than one photon at a time --- doesn't contribute to photon loss. This kind of optical coupling from the waveguide into the PD absorber and back into the waveguide can be accomplished by using a vertical directional coupler design discussed in detail in \sec{wg}.

\begin{figure}[h!]
 \begin{center}
\begin{overpic}[width=0.35\textwidth]{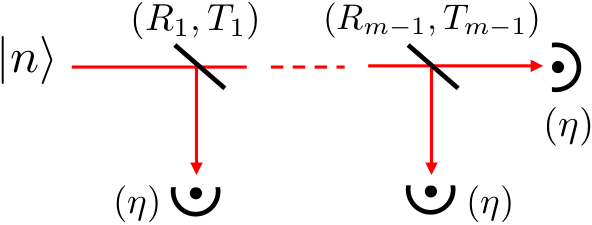}
\put(29, -5){$a'_1\ (n_1)$}
  \put(64, -5){$a'_{m-1}\ (n_{m-1})$}
 \put(104, 26){$a'_m\ (n_m)$}
\end{overpic}
\end{center}
\caption{Model of a PNR segmented photodetector with $R_{j}+T_{j}\equiv r_{j}^{2}+t_{j}^{2}=1, \forall j\in[1,m]$. \label{fig:seg}}
\end{figure}

In order to further simplify the model for \sec q, at no cost to its generality, we can recast our segmented detector as the SPAD sequence depicted in Fig.~\ref{fig:seg}. 
In that case, the loss channel corresponding to deviations to Eq.~\eqref l becomes equivalent to $\sqrt{1-\eta}$, where $\eta$ is the quantum efficiency of the terminally coupled SPADs in Fig.~\ref{fig:seg}. While there is no fundamental difference between radiative losses in Fig.~\ref{waveguide} and Fig.~\ref{fig:detail} and $\eta<1$ in Fig.~\ref{fig:seg}, there is, again, a conceptual difference between $\alpha^{2}<1$, which doesn't lead to photon loss since the photon can re-enter the waveguide, and $\eta<1$, which does constitute photon loss. In addition, $\eta<1$ also accounts for the mechanism by which the photon can be absorbed in a SPAD without causing an avalanche. Our theoretical model in \sec q will also account for dark counts and electrical cross-talk. 

\section{Segmented waveguide detector design}\label{sec:wg}

\begin{figure}
 \includegraphics[width=0.5\textwidth]{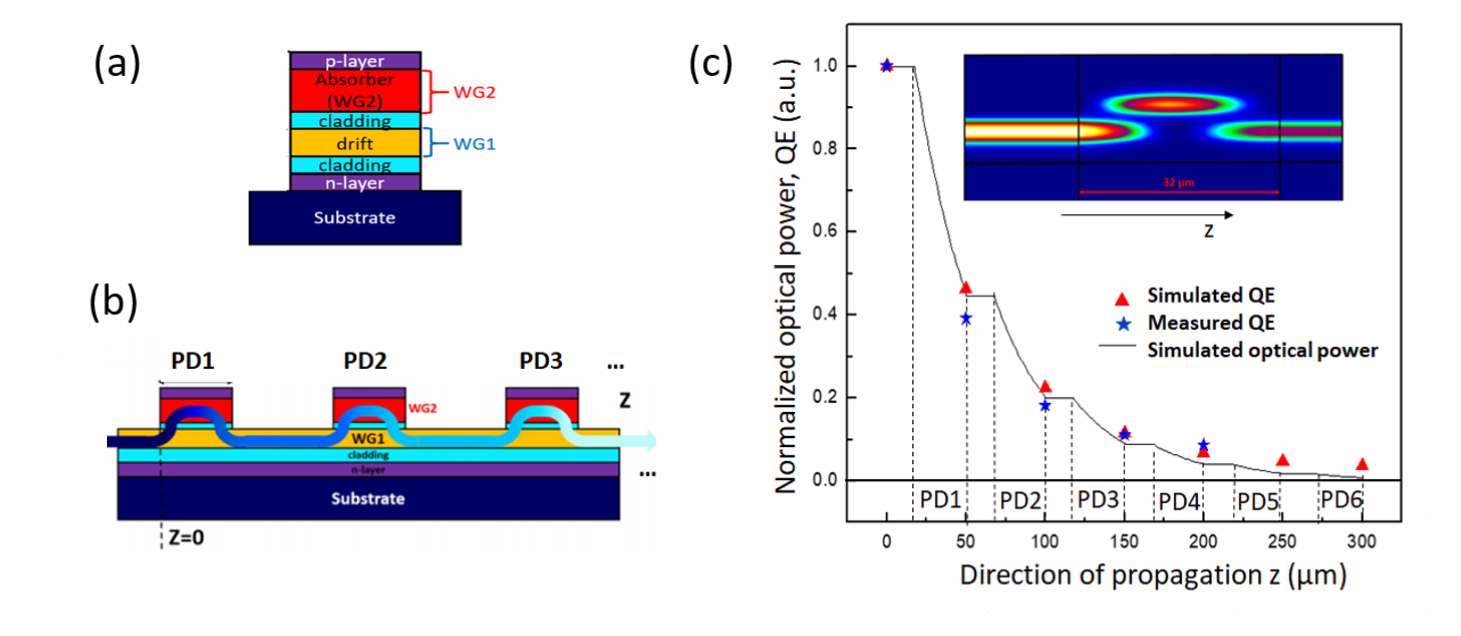}
  \caption{(a) Cross-section of waveguide photodiode~\cite{Yu:18}; (b) Side view schematic of light propagating in the segmented waveguide photodetector. (c) Normalized optical power and QE of PD1 to PD6 in the segmented photodetector. The inset shows the optical intensity in PD1 with a PD length of 32$\mu$m.}{\label{A_fig1}} 
 \end{figure}

To verify the optical design of the segmented detector, we previously reported a monolithically integrated InP-based p-i-n waveguide photodetector  consisting  of  6 PIN photodiodes (PDs),    coupled  to  one  waveguide~\cite{Yu:18}, Fig.~\ref{A_fig1}(a).  Optical coupling from the waveguide into the PD absorber and back into the waveguide was accomplished by using a vertical directional coupler design as shown in Fig.~\ref{A_fig1}(b).
 Input light propagates in the passive waveguide WG1 and couples into the absorption waveguide (WG2) of the PD where electron-hole pairs are generated. Residual light in the WG2 couples back into WG1 at the end surface of each PD. 
 
\begin{figure}
\includegraphics[width=0.5\textwidth]{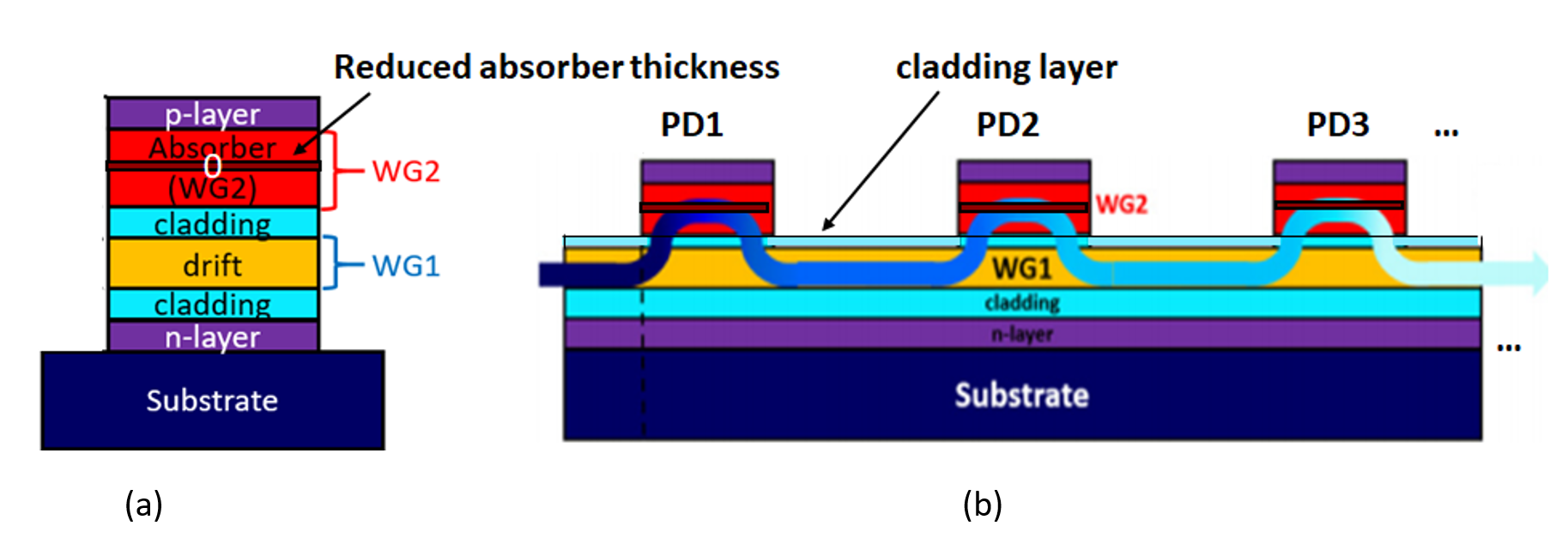}
\caption{(a) Cross-section of of new waveguide photodiode design with thin absorber; (b) Side view schematic of light propagating in the segmented waveguide photodetector with cladding layer.}{\label{A_fig2}} 
\end{figure}

 By using a segmented photodetector with six elements, we experimentally demonstrated an overall quantum efficiency (QE) of 90$\%$. Fig.~\ref{A_fig1}(c) shows the simulated optical power along the segmented photodetector for the design in \cite{Yu:18} along with the QE of each PD. Each period in the photodetector was 50 $\mu$m long with a 32 $\mu$m-long PD and a 18 $\mu$m-long passive waveguide. The solid line shows how the optical power decreases while propagating in direction z. We simulated a total optical loss of 1$\%$ at the front and rear side of each PD. WG1 was assumed to be lossless.  The red and blue symbols show the simulated and measured QE for each individual PD in the photodetector. Here, the optical power was referred to the input power in the waveguide at z = 0. The simulated total QE was 96.5$\%$ which was close to the measured value of (90$ \pm 5)\%$ \cite{Yu:18}. The error  originated from uncertainty in determining the fiber-chip coupling loss; the difference between simulation and measurement can be explained by fabrication tolerances and non-zero waveguide loss. 
To reduce the latter, the waveguide length can be reduced to < 10$\mu$m. To further reduce the radiation loss and enable segmented detectors with larger PD count we designed a new structure and made two changes comapred to \cite{Yu:18}: (i) we added an additional cladding layer on top of the passive waveguide WG1, and (ii), we also reduced the thickness of the absorption layer from 30 nm in \cite{Yu:18} to only 6 nm as seen in Fig.~\ref{A_fig2}.

\begin{figure}
 \includegraphics[width=0.5\textwidth]{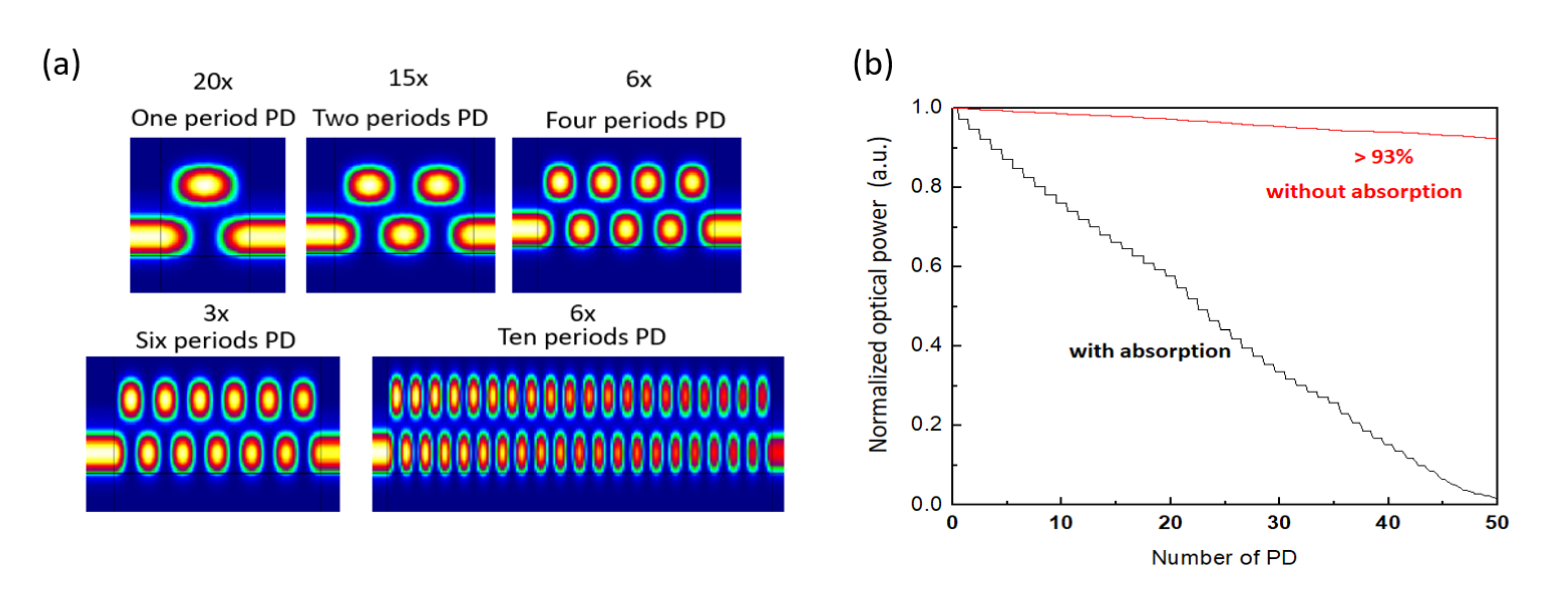}
  \caption{(a) Simulated optical intensity in PDs with various lengths; (b) Normalized optical power with (black) and without (red) absorption in the 50-element segmented detector with 8 mm total length.}{\label{A_fig3}} 
\end{figure}

A small active volume is beneficial since it helps reducing the dark current, jitter, and increases the PD's count rate. Given the fact that the PD length can only be an integral multiple of the mode beat length L~\cite{Wang2017}, we designed a 50-element segmented detector with 20 PDs with PD length L, followed by 15 PDs with PD length 2L, 6 PDs with 4L,  3 PDs with 6L, and 6 PDs with 10L as demonstrated in Fig.~\ref{A_fig3}(a). This ensures complete absorption and a similar number of photogenerated electron-hole pairs in each of the 50 PDs.  
Fig.~\ref{A_fig3}(b) shows the stepwise decay of the simulated optical power in the segmented detector with uniform PD QE of 2.5\% in each PD. We estimated the overall radiation loss to be as low as 7\% by simulating the same structure without including any imaginary indices, Fig.~\ref{A_fig3}(b). It should be mentioned that additional loss originating from WG1 can be as low as $1 \%$ or  $4 \%$ assuming  either a low-loss $\rm Si_{3}N_{4}$ waveguide (0.1 dB/cm \cite{munoz2019foundry}) or an InGaAsP waveguide (0.4 dB/cm \cite{d2015low}). 

\section{Quantum modeling of the segmented detector}\label{sec:q}

\subsection{Positive-operator-valued measure (POVM)}

The most general measurements in quantum physics are known as positive-operator-valued measures (POVMs)~\cite{Helstrom1976} and have been widely used to model photodetection~\cite{vanEnk2017a, vanEnk2017b}.

Each measurement outcome $k$ is given by a Hermitian operator, the POVM element ${\Pi}_k$,  with nonnegative eigenvalues such that the probability of an outcome for a quantum density operator ${\rho}$ is given by
\begin{equation}
p_k = \text{Tr}(\rho {\Pi}_k).
\end{equation}
These operators satisfy the completeness property, $\sum_{k} {\Pi} _k = I$ which makes the sum of probabilities for different outcomes $\sum_{k} p_k = 1$. Therefore, these POVM elements completely describe all possible measurement outcomes. For a phase insensitive detector the POVM element for $k$ independent SPAD pulses, a.k.a.\ ``clicks,'' is given by
\begin{equation}
\Pi _k = \sum_{n = 0}^{\infty} P(k|n) |n\rangle \langle n|,
\label{eq:POVM_general}
\end{equation}
where $P(k|n)$ is the conditional probability of getting $k$ clicks given an $n$-photon input. Unlike projective, von Neumann measurements, the POVM elements are not orthogonal measurements and we have, in general, 
\begin{equation}\label{povm}
{\Pi}_k  {\Pi}_{k'} \neq \delta_{kk'}\Pi_k.
\end{equation}
Note that orthogonal POVM elements are projective measurements and we can define the purity of the POVM element for outcome $k$ as
\begin{equation}\label{purity}
\text{Purity}(\Pi _k) =  \frac{\text{Tr}(\Pi _k^2)}{{ \text{Tr}(\Pi _k)^2 }}.
\end{equation}

The purity is obviously a positive quantity. It also satisfies
\begin{equation}
\frac1D\leq \text{Purity}(\Pi_k) \leq 1,
\end{equation}
where $D$ is the dimension of the Hilbert space associated with POVM element $\Pi_k$, i.e., the number of projectors with nonzero probabilities in the sum of Eq.~(\ref{eq:POVM_general}). Clearly, a POVM element giving a completely random outcome will yield $D\to\infty$ and a purity of zero. Therefore, the reciprocal of the purity Purity$(\Pi_k)^{-1}$ can be seen intuitively as a loose estimator of the number of input states $|n\rangle$ that result in click numbers $k$ with significant probabilities.

Experimental quantum-detector tomography has been achieved in a number of different situations~\cite{Lundeen2009NP,Piacentini:15,1367-2630-17-10-103044}. Here we study the theoretical purity of the POVM elements of Eq.~\eqref{purity} versus the input photon number, in the presence of nonidealities such as loss channels, which make $k<n$ due to radiative losses, detector absorption without avalanche, etc., and such as electrical cross-talk and dark counts, which make $k>n$ due to non-photon-triggered avalanches,  absorption-triggered parasitic flashes  on the detector surface, and afterpulsing. Note that afterpulsing, i.e., dark counts caused by charges from previous avalanches trapped in impurity levels,  is also conditioned on the number of clicks on a given SPAD but we'll treat this effect as a second order one and neglect it, effectively treating afterpulsing as dark counts.

\section{Model}
\subsection{General model of a segmented detector
}
We model Fig.~\ref{fig:seg} by using a Heisenberg picture approach.  The quantum input mode is described by annihilation operator $a_1$ and input Fock state $|n\rangle$,  the other $m-1$ input modes $a_2$,  $a_3$, ..., $a_{m}$ are vacuum modes, and the detection  modes are $a'_1$,  $a'_2$, ..., $a'_{m}$. We consider $m-1$  
beamsplitters $(T_{j}, R_{j})$ and $\eta=1$ (no losses) for all modes.  The input quantum state is
 \begin{equation}
 |n\rangle = \frac{{a^{\dagger}_1}^n}{\sqrt{n!}}|0\rangle^{\otimes^{m}}.
 \end{equation}

In order to find the probability of an outcome it is convenient to write the output quantum state in terms of detection modes. Using backpropagation of the detection modes, $a'_i$s we get
\begin{equation}\label{Eqn}
\begin{aligned}
a^{\dagger}_1 =  r_{1}a_1'^\dagger  + \sum_{k = 2}^{m-1}\big[ \prod_{l = 1}^{m-1} t_l\big]r_{k-1}a_k'^\dagger  + \prod_{l = 1}^{m-1}t_la_m'^\dagger, 
\end{aligned}
\end{equation}
and the output quantum state is 
\begin{widetext}
\begin{align}
|\psi\rangle_{out} 
& =\frac{1}{\sqrt{n!}}{\left(  r_{1}a_1'^\dagger + \sum_{k = 2}^{m-1}
\prod_{l = 1}^{m-1} t_l
\,r_{k-1}\,a_k'^\dagger + \prod_{l = 1}^{m-1}t_l\,a_m'^\dagger  \right)^n} |0\rangle^{\otimes^{m}}
\end{align}
and the multinomial expansion yields
\begin{equation}  
%\begin{aligned}
|\psi\rangle_{out}  = \underbrace{\sum_{n_1=0}^n\cdots\sum_{n_m=0}^n}_{\sum_{j=1}^{m}n_j=n} 
\frac{\sqrt{n!}}{n_1!n_2!...n_m!}r_1^{n_1}\,\prod_{k=2}^{m-1}\tau_{1,k-1}^{n_k} r_k^{n_k}\,\tau_{m-1,1}^{n_m} \prod_{i = 1}^{m}\,a_i'^{\dagger n_i}
|0\rangle^{\otimes^{m}},
%\end{aligned}
\end{equation}
%\end{widetext}
where each $n_i$ can take any value from 0 to $n$ and $\tau_{i,j} = t_i\dots t_j$. 

Given $n$ input photons,  the probability $P_m(k|n)$ of getting $k$  clicks from $m$ SPADs --- where each click may result from one or several simultaneous photons, since single SPADs aren't PNR --- is, in the lossless case,  
%\begin{widetext}
\begin{align}\label{Pkn}
P_m(k|n)&=  \underbrace{\sum_{n_1=0}^n\cdots\sum_{n_m=0}^n}_{(*)\sum_{i=1}^{k}n_{i}=n} \frac{n!}{\prod_{i=1}^{m} n_i!}\ X,\\
\text{where }X &= \left(r_1^{n_1}\prod_{k=2}^{m-1}\tau_{1,k-1}^{n_k} r_k^{n_k}\,\tau_{m-1,1}^{n_m}\right)^2 \\
 &=R_1^{n_1}T_1^{n-n_1}\times R_2^{n_2}T_2^{n-n_1-n_2}
 \times\cdots\times R_{m-1}^{n_{m-1}}T_{m-1}^{n-\sum_{j=1}^{m-1} n_j}.
\end{align}
\end{widetext}
The asterisk in Eq.~\eqref{Pkn} symbolizes the following constraint: in this lossless case, $k$ clicks will be obtained if and only if $k$ different SPADs out of $m$ receive {\em at least} one photon, and the other $m-k$ SPADs receive zero photon. A fully explicit formula will be given for the symmetrized detector in the next section.

Note that our goal is to actually split the $n$ input photons among $m\gg n$ modes with never more than 1 photon per mode. The first beamsplitter's reflectivity must then be much less than $n^{-1}$.  Taking all beamsplitters identical is clearly not optimal since the subsequent modes will gradually see fewer photons and can therefore afford larger reflectivities without running the to risk of detecting more than one photon. Also, at the end of the segmented device, the last beam splitter should clearly be balanced since the constraint has to be symmetric for both its output ports. Bearing all this in mind, a symmetrized device appears to be the optimal choice. We investigate it next. 

\subsection{Symmetrized segmented detector}
\subsubsection{Lossless case}
\begin{figure}[h!]
        \includegraphics[width=0.5\textwidth]{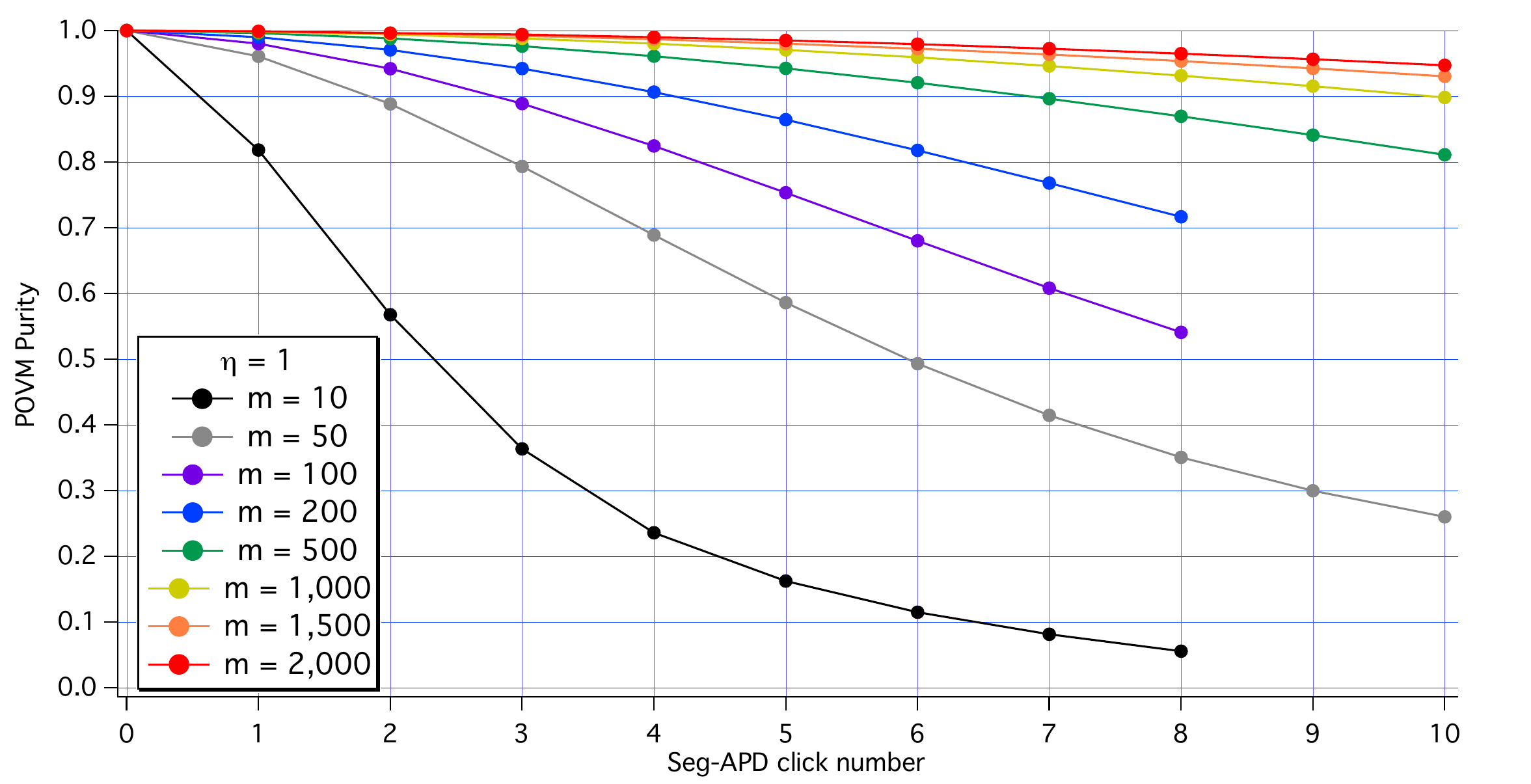}
        \caption{POVM element $\text{Purity}(\prod_{k})$ versus click number $k$,  for different  $m$.}{\label{pur_eta_1}}
\end{figure}

We take $\eta=1$ and the beamsplitters' reflectivities to be  $R_{j}=\frac{1}{m-j+1}$,  where $ j\in[1,m-1]$,  yields the simplification
\begin{equation}
X =  \frac{1}{m^n},
\end{equation}
and makes the symmetrized segemented detector equivalent to a symmetric beamsplitter tree
, to the notable difference that SPADs aren't terminally coupled here and so their nonideal QE doesn't contribute to detection losses. Eq.~\eqref{Pkn} thus simplifies to 
\begin{equation}\label{eq_prob}
P_m(k|n) =  \frac{n!}{m^n}  {m\choose k} \underbrace{\sum_{n_1=1}^n\cdots\sum_{n_k=1}^n}_{\sum_{j=1}^{k}n_j=n} \frac{1}{\prod_{i=1}^{k} n_i!}.
\end{equation}      
\begin{figure}[h!]
\begin{center}
 \includegraphics[width=0.5\textwidth]{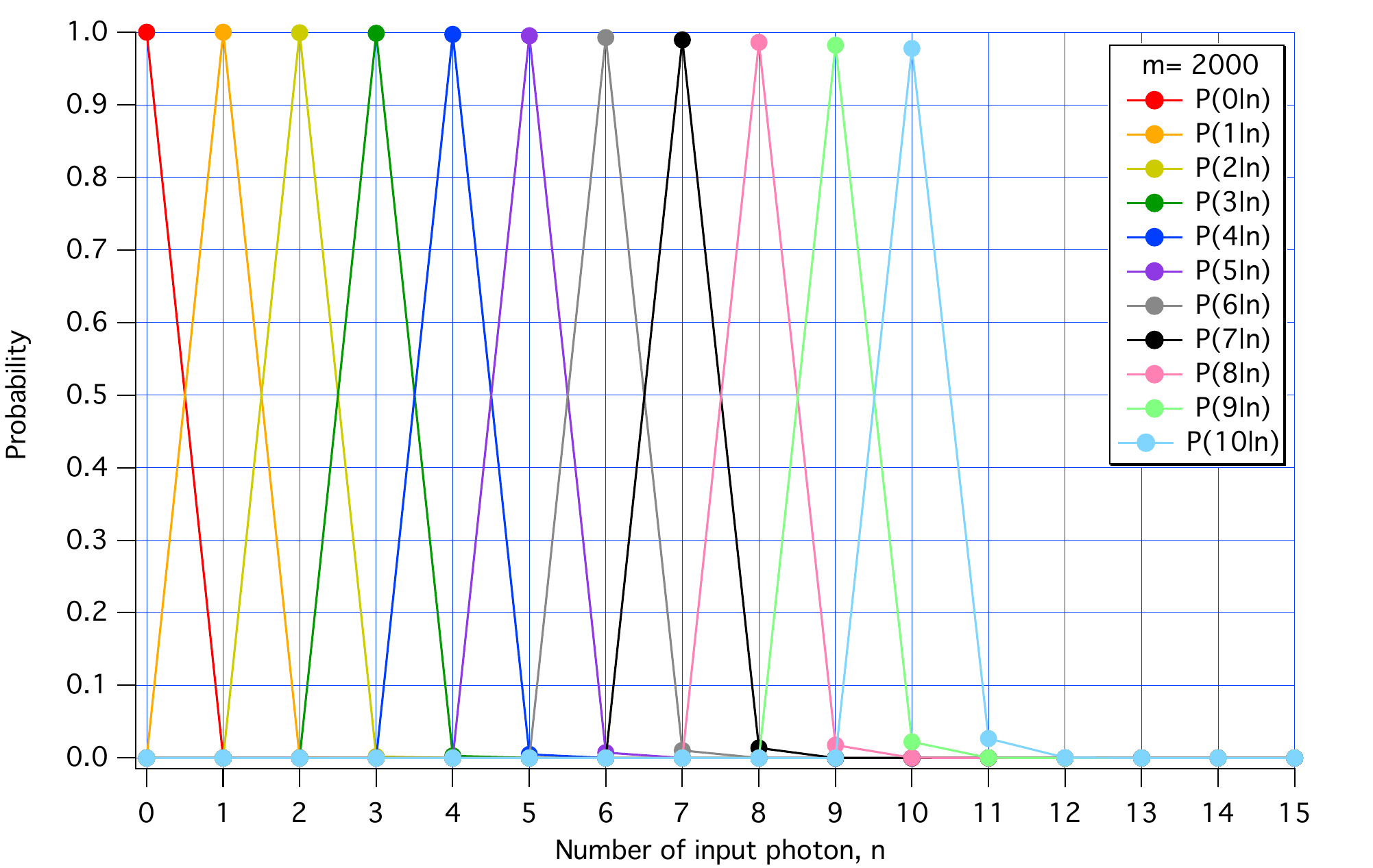}
  \caption{Conditional probabilities $P_{2000}(k|n)$ versus $n$, for $\eta = 1$.}{\label{prob_eta_1}} 
 \end{center}
 \end{figure}
These conditional probabilities are plotted in Fig.~\ref{prob_eta_1} for $m=2000$. Unsurprisingly, they increase for larger photon number $n$ as the SPAD number $m$ increases. This translates directly into the POVM purities, plotted in Fig.~\ref{pur_eta_1}.
As can be seen in that figure, reasonably good PNR performance (POVM purity of at least 90\%) is reached for $n\sim10$ with $m\sim1000$.

\subsubsection{Lossy case}
\begin{figure}[t!]
  \includegraphics[width=0.5\textwidth]{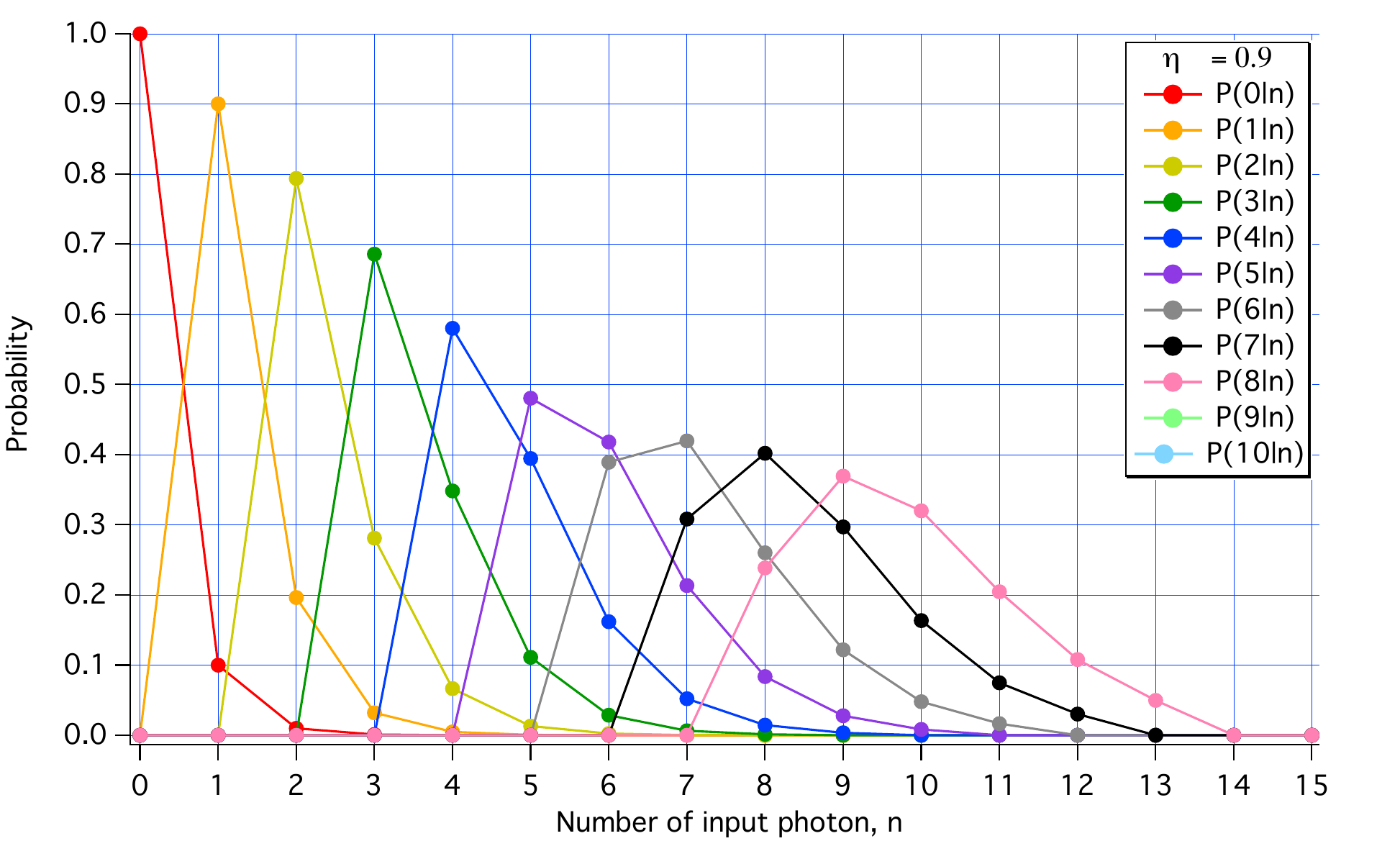}
  \caption{Conditional probabilities $P_{50}(k|n)$ versus $n$, for $\eta = 0.9$.}\label{prob_eta_0.9}
 \end{figure}

We now consider the effect of photon losses in each detection mode, i.e., $\eta < 1$. Again, $\eta$ should not be misconstrued to be the SPAD, which plays no role in photon losses.
Again, $\eta$ should not be misconstrued as the SPAD absorption efficiency because the latter plays no role in photon losses, as unabsorbed photons can enter back into the waveguide and get detected by subsequent SPADs. Further more, the quantity $1-\eta$ is the probability of a photon exiting the waveguide undetected or a photon was absorbed without causing an avalanche, losing its chance for further detection. We assume that the parameter $\eta$ is independent of the photon number. 
The probability to get zero clicks in one mode is 
\begin{equation}\label{eq:0}
P_{1}(0|n, \eta) = (1-\eta)^n.
\end{equation}
Likewise, the probability to get one click in one mode is ~\cite{Scully}
\begin{equation}
P_{1}(1|n, \eta) =\sum_{k=1}^{n} {{n}\choose{k}}  {\eta}^k{(1-\eta)^{n-k}}
=1-(1-\eta)^n.
\end{equation}
It is important to note that in the sum over $k$ starts with 1 here because we neglected dark counts. Therefore, Eq.~\eqref{eq_prob} can be generalized to the lossy case as
\begin{widetext}
\begin{equation}
\begin{aligned}\label{loss_prob}
P_{m}(k|n, \eta) = n!\left(\frac{1-\eta}{m}\right)^n {{m}\choose{k}} \underbrace{\sum_{n_1=0}^n \cdots\sum_{n_m=0}^n}_{\sum_{j=1}^{m}n_j=n} 
\frac{1}{\prod_{j=1}^{m} n_j!}  
 \prod_{l=n_1}^{n_k}\left[\left(\frac1{1-\eta}\right)^{l}-1\right],
\end{aligned}
\end{equation}
\end{widetext}
In Eq.~\eqref{loss_prob}, in addition to multinomial coefficient for probability we also have the product term due to non-unity quantum efficiency. It is worth noting that the probability of getting zero clicks is still the same as the case of one detector, i.e., Eq.~\eqref{eq:0}, and for $\eta$ = 1, Eq.~\eqref{loss_prob} turns out to be same as Eq.~\eqref{eq_prob} for $n_i \geq 1$  where $i$  $\in$ $[1, k]$ for k clicks.  For  $\eta<1$ the computer simulations run extremely slowly for higher values of $m$ (reminiscent of the boson sampling problem), therefore we were limited to $m = 50$ for the calculation of conditional probabilities at $\eta = 0.9, 0.99, 0.999$. The results at $\eta=0.9$ are displayed, for illustrative purposes, in Fig.~\ref{prob_eta_0.9}.
The degradation of the count probability with photon loss is evident, compared to Fig.~\ref{prob_eta_1}. Also recall that $\eta=0.9$ means 10\% loss per detection mode which is a very poor performance as previous experimental work on low loss waveguides shows one can do much better~\cite{Bauters2011}. 

\begin{figure}[h!] 
        \includegraphics[width=0.5\textwidth]{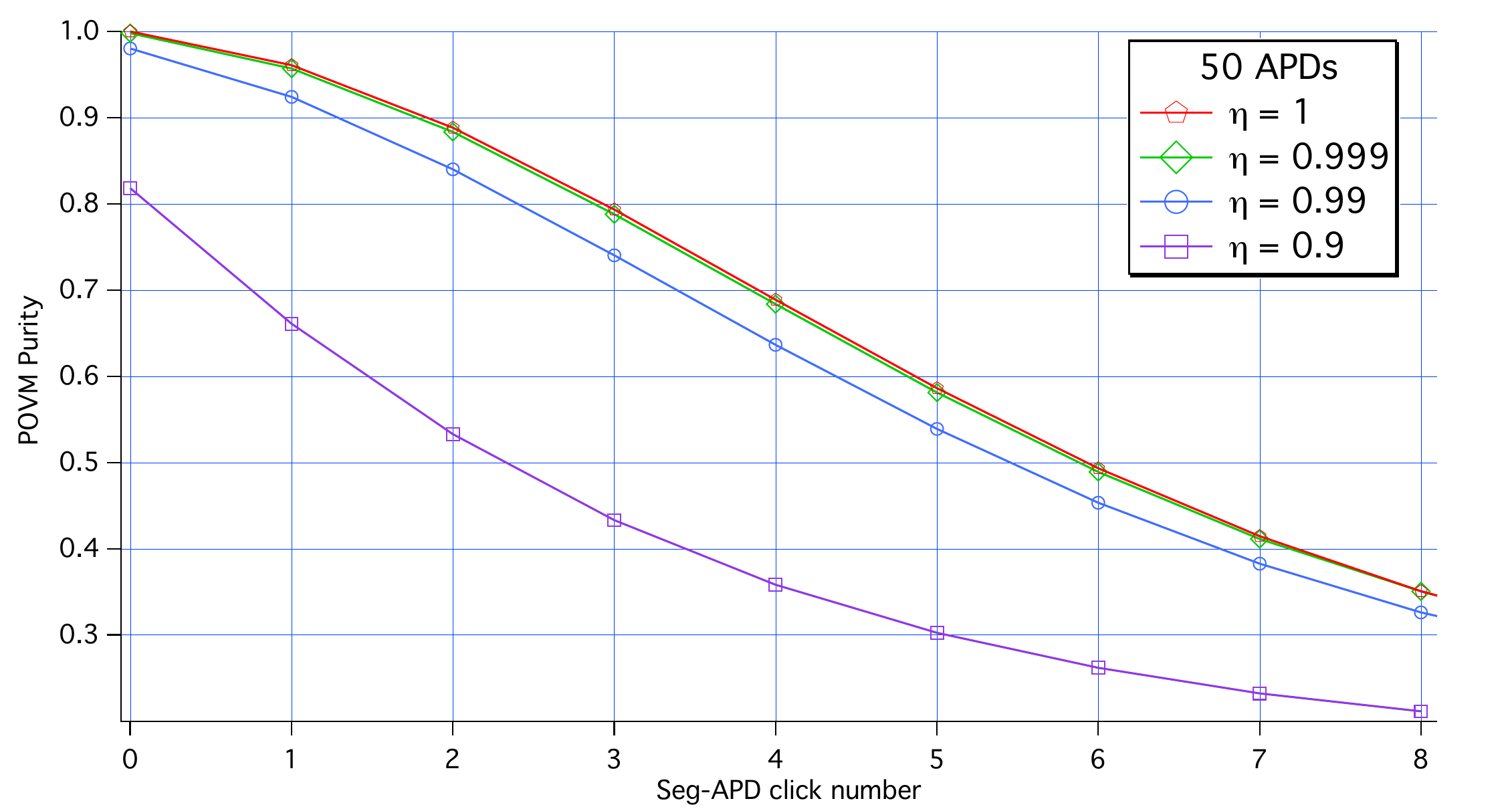}
        \caption{POVM element $\text{Purity}(\prod_{k})$ versus click number $k$, for several values of $\eta$ at $m = 50$}{\label{pur_eta_0.9}}
        \end{figure}

The purity calculation, displayed in Fig.~\ref{pur_eta_0.9}, is particularly illuminating.
Indeed, it is clear that, as $\eta$ increases beyond the low $\eta=0.9$ level, the photon losses have a decreasing to negligible ($\eta=0.999$, i.e., 0.1\% loss per detector) effect on purity, which is essentially limited by $m$, as per Fig.~\ref{pur_eta_1}. This is an interesting result. It is likely that the same level of photon loss may have a more detrimental effect as $m$ increases, however, the exact scaling of this effect is not yet known, due to the long computation times for the nonideal case.
{Figures \ref{pur_eta_0.9} and \ref{prob_eta_0.9} can be related by the aforementioned intuitive meaning of the POVM purity:  
consider, for example, $k=5$ in Fig.~\ref{pur_eta_0.9}, for which the POVM purity $\simeq$ 0.3. This implies that the number of input states leading to clicks with significant probabilities is about 3 to 4, which corresponds to the number of points making up most of the purple $P(5|n)$ peak in  Fig.~\ref{prob_eta_0.9}. Also, one can clearly see that, as the purity decreases with $k$ in Fig.~\ref{pur_eta_0.9}, the conditional probabilities have broader and broader supports with larger $k$ in Fig.~\ref{prob_eta_0.9}.}
\subsubsection{Dark count noise modeling}
\begin{figure}[h!] 
 \begin{center}
        \includegraphics[width=0.5\textwidth]{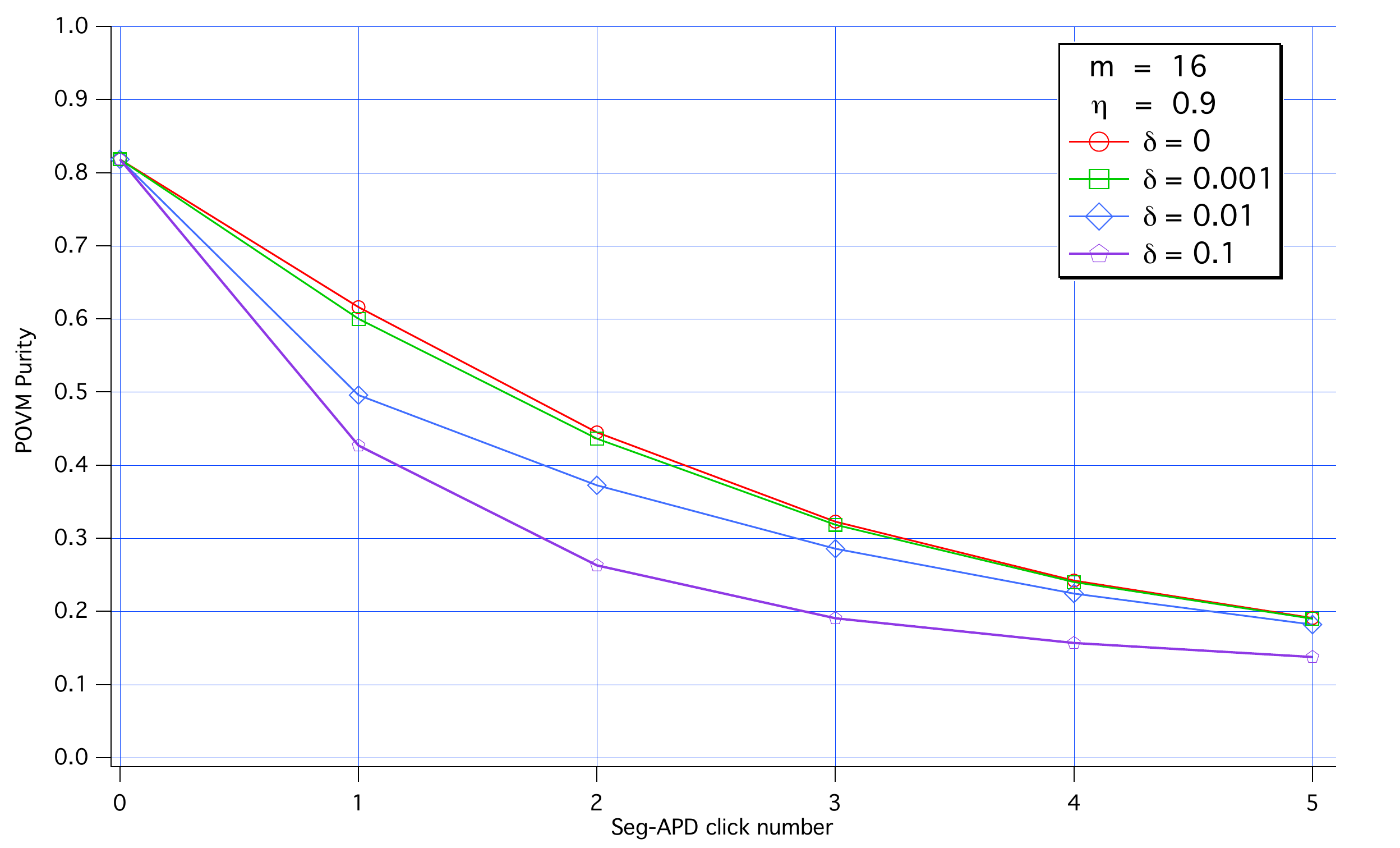}
        \caption{POVM element $\text{Purity}(\prod_{k})$ versus click number $k$, for several values of dark count probabilities, $\delta$ at $m = 16$}{\label{fig:Dark_count}}
         \end{center}
        \end{figure}

We now model dark counts for the segmented detector. We start with a fixed dark count probability, $\delta$, independent of the input photon-number. Recall that a non-PNR detector has 2 POVM elements $\Pi_1^d$ and $\Pi_0^d$ corresponding to 2 detection events, click and no-click respectively.  In the presence of dark counts, the probability of having no click becomes, from Eq.~\eqref{eq:0},
\begin{equation}
       P_0^d = (1-\delta)(1-\eta)^n,
       \label{dark_zero}
\end{equation}
which can be interpreted as the joint probability of Eq.~\eqref{eq:0} (no photon detected from the incident light, with probability  $(1-\eta)^n$) and no click from dark counts, with probability of $(1-\delta)$. Since both of those events are independent, the overall probability of them occurring simultaneously is the multiplication of individual probabilities. Thus,  the probability of registering a click is
\begin{equation}
P_1^d = 1-P_0^d = 1-(1-\delta)(1-\eta)^n,
\label{dark_one}
\end{equation}  
and  the 1-click POVM for a phase insensitive detector is
\begin{equation}
  \Pi_1^d = \sum_{n=0}^{n=\infty}\left[1-(1-\delta)(1-\eta)^n\right]|n\rangle\langle n|. 
\label{eq:click_povm}
\end{equation}
Note that the sum in Eq.~\eqref{eq:click_povm} is now starting from zero which accounts for the possibility of dark counts, of probability $\delta$,  with no light incident on the detector. The $k$-click POVM  is then
\begin{widetext}
\begin{equation}
\begin{aligned}
P_{m}(k|n, \eta) = n!\textcolor{black}{(1-\delta)^m} \left(\frac{1-\eta}{m}\right)^n {{m}\choose{k}} \underbrace{\sum_{n_1=0}^n \cdots\sum_{n_m=0}^n}_{\sum_{j=1}^{m}n_j=n}
\frac{1}{\prod_{j=1}^{m} n_j!}  
 \prod_{l=n_1}^{n_k}\left[\textcolor{black}{\frac{1}{{1-\delta}}}\left(\frac1{1-\eta}\right)^{l}-1\right].  
\end{aligned}
\end{equation}
\end{widetext}
Fig.~\ref{fig:Dark_count} displays the simulated POVM purities for $m = 16$, $k\leqslant5$, $\eta = 0.90$, 
and dark count probabilities $\delta$ = 0, 0.001, 0.01, and 0.1. (Due to the heavy numerical load, we could not compute for larger SPAD numbers $m$.) We see that the POVM purities decrease as dark count probability increases, unsurprisingly. However, the key point is that $\delta$ = 0.1\% is practically indistinguishable from zero dark counts in this case, where the scalability of the segmented detector and efficiency $\eta$ are the main limitation to PNR operation.

\subsubsection{Cross-talk noise modeling}
\begin{figure}
    \centering
    \includegraphics[width=.5\textwidth]{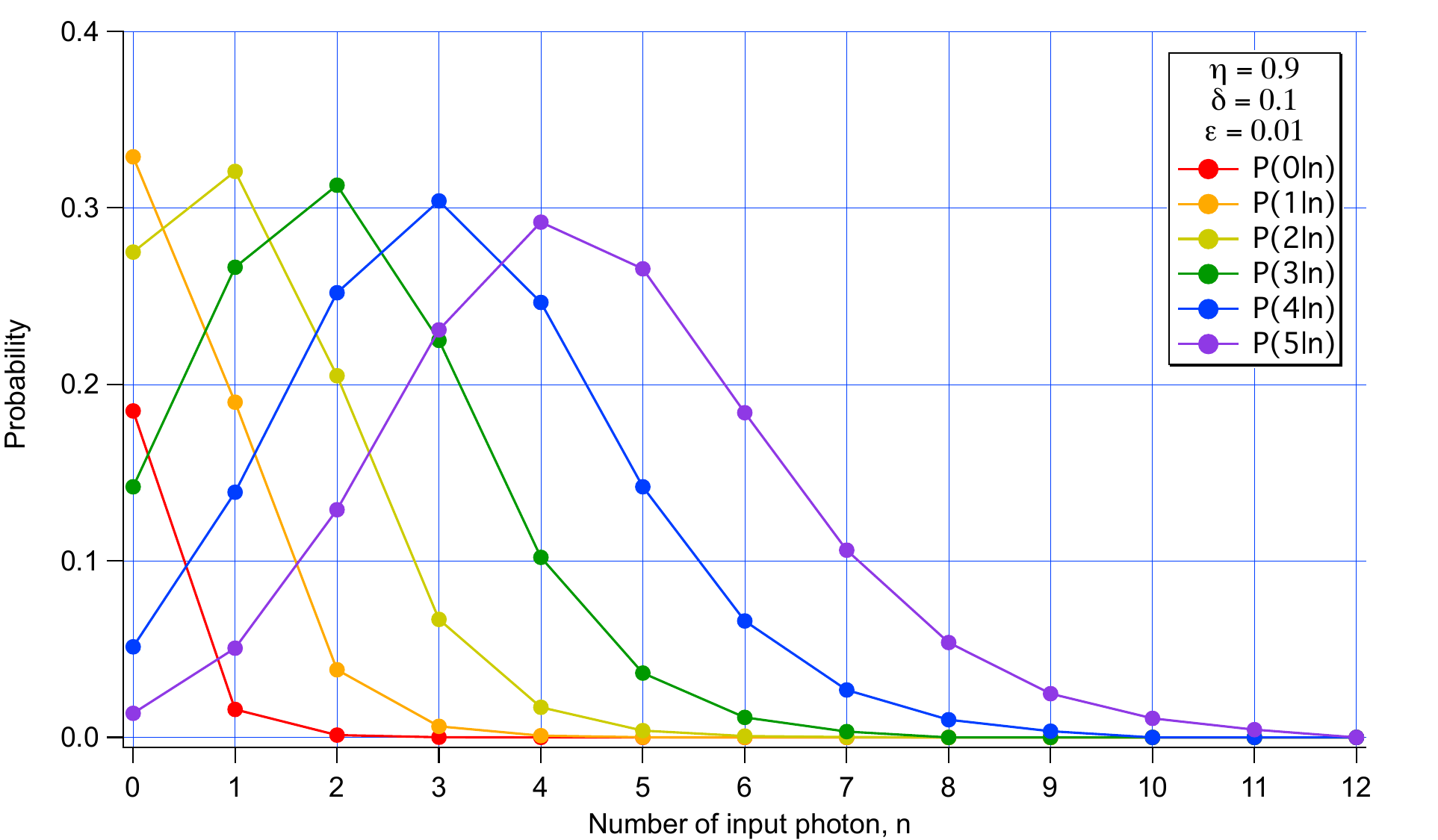}
    \caption{Conditional probabilities $P_{16}(k|n)$ versus $n$, for $\eta = 0.9$, $\delta = 0.1$, and $\epsilon = 0.01$. }
    \label{fig:my_new_prob}
\end{figure}

A cross-talk event is registered when an avalanche in a particular SPAD causes an avalanche in the neighboring SPADs~\cite{Rech2008}. Since the dark-count rate can be extremely low in our design, we only consider the cross-talk events due to the incident light.  The effect of cross-talk can be reduced by increasing the distance between consecutive SPADs but this will increase the propagation losses in the waveguide. Thus, it becomes critical to account for registered clicks caused due to cross-talk. To model the cross-talk, we first consider the simplest case of two SPADs on the waveguide. In this case, the POVM set has 3 elements corresponding to zero-, one-, and two-click detection outcomes. If $n$ photons are coupled to the waveguide, $n_1$ photons are coupled to first SPAD and $n_2$ are coupled to the second SPAD. We define a new parameter $\epsilon$, which is the probability cross-talk event caused by an avalanche in neighboring SPADs. In the two-SPAD case, the probability of a cross-talk event registered by the first SPAD is $(1-\epsilon)^{n_2}$, where $n_2 = n-n_1$ is the number of photons passed to second SPAD. Thus, the probability of getting zero-click outcome is 
\begin{widetext}
\begin{align}
P^{d,\epsilon}_0 &= \left(\frac{1}{2}\right)^n\underbrace{\sum_{n_1=0}^n\sum_{n_2=0}^n}_{\sum_{j=1}^{2}n_j=n}\frac{n!}{n_1!n_2!}\underbrace{(1-\delta)(1-\eta)^{n_1}(1-\epsilon)^{n_2}}_{\text{*}}\underbrace{(1-\delta)(1-\eta)^{n_2}(1-\epsilon)^{n_1}}_{\text{\#}}\\
& = (1-\delta)^2(1-\eta)^n (1-\epsilon)^n,
\end{align}
%\end{widetext}
where `$*$' and `$\#$' are the probabilities of having no click from the first and second SPAD respectively. Likewise, for one- and two-click detection events we have 
%\begin{widetext}
\begin{align}
P^{d,\epsilon}_1 &= \left(\frac{1}{2}\right)^n\underbrace{\sum_{n_1=0}^n\sum_{n_2=0}^n}_{\sum_{j=1}^{2}n_j=n}\frac{n!}{n_1!n_2!}\{\underbrace{[1-(1-\delta)(1-\eta)^{n_1}(1-\epsilon)^{n_2}]}_{\text{First SPAD clicked}}
\underbrace{[(1-\delta)(1-\eta)^{n_2}(1-\epsilon)^{n_1}] }_{\text{No click from second SPAD}} \nonumber \\
&+ \underbrace{[1-(1-\delta)(1-\eta)^{n_2}(1-\epsilon)^{n_1}][(1-\delta)(1-\eta)^{n_1}(1-\epsilon)^n_2]}_{\text{The other way around}}\}, 
\end{align}
\begin{equation}
P^{d,\epsilon}_2 = \left(\frac{1}{2}\right)^n\underbrace{\sum_{n_1=0}^n\sum_{n_2=0}^n}_{\sum_{j=1}^{2}n_j=n}\frac{n!}{n_1!n_2!}[1-(1-\delta)(1-\eta)^{n_1}(1-\epsilon)^{n_2}] [(1-(1-\delta)(1-\eta)^{n_2}(1-\epsilon)^{n_1}].
\end{equation}
We can generalize to $m$ detectors and $k$ clicks as
\begin{equation}
P_{m}(k|n, \eta, \delta, \epsilon) = C n! \left(\frac{1-\eta}{m}\right)^n {{m}\choose{k}} \underbrace{\sum_{n_1=0}^n \cdots\sum_{n_m=0}^n}_{\sum_{j=1}^{m}n_j=n}\left\{ \frac{1}{\prod_{j=1}^{m} n_j!}  
 \prod_{l=n_1}^{n_k}\left[{\frac{1}{{(1-\delta)(1-\epsilon)^n}}}\left(\frac{1-\epsilon}{1-\eta}\right)^{l}-1\right] \right\}, 
\end{equation}
\end{widetext}
where $C = (1-\delta)^{m} (1-\epsilon)^{(m-1)n}$ is a constant depending on dark count and cross-talk rates. 
In Fig.~\ref{fig:my_new_prob}, we plot these conditional probabilities for $m = 16$ with the dark-count probability $\delta=0.9$ and cross-talk $\epsilon = 0.01$. It can be clearly seen that in the presence of dark-count and cross-talk events, the peaks are broadened and shifted to the left in comparison to Fig.~\ref{prob_eta_0.9}. This implies that a $k$ click event could happen even if less than $n=k$ photons are incident to the detector, which is caused by registered avalanches due to dark-count and cross-talk. In addition, the probability of getting zero click reduces substantially as seen in red curve in Fig.~\ref{fig:my_new_prob}.

\begin{figure}
    \centering
    \includegraphics[width=.5\textwidth]{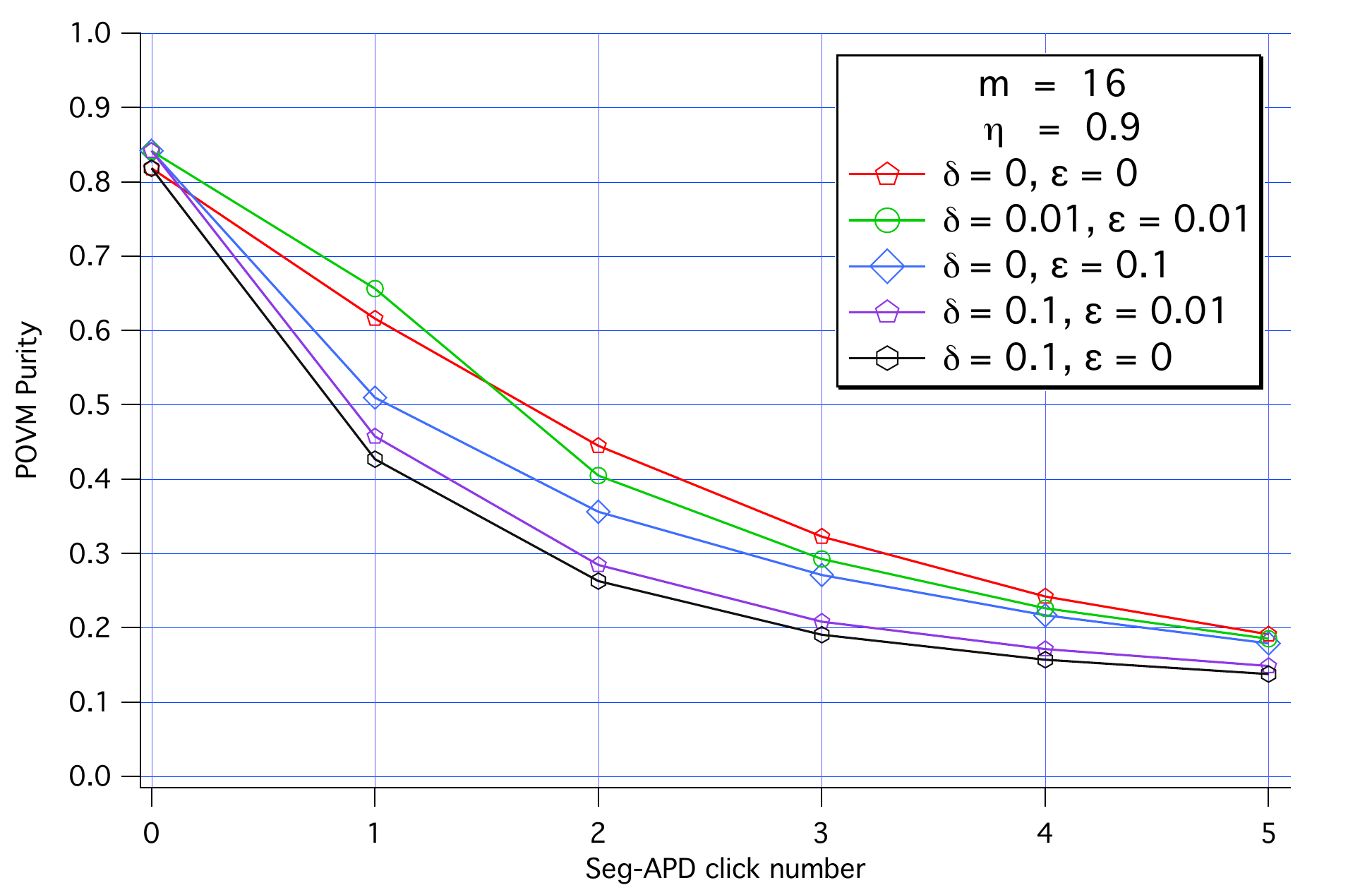}
    \caption{POVM element $\text{Purity}(\Pi_{k})$ versus click number $k$, for several values of $\delta$ and $\epsilon$ at $m = 16$.}
    \label{fig:my_fig_purity}
\end{figure}

{
In Fig.~\ref{fig:my_fig_purity}, we plot in the POVM purity for $\eta = 0.9$ and $m = 16$ for $\delta = 0, 0.01, 0.1$ and $\epsilon = 0, 0.1, 0.01$. The general trend shows that the POVM purity decreases as dark-count and cross-talk rates increase, unsurprisingly. Note the slight increase in the POVM purity for $\delta = 0.01$ and $\epsilon = 0.01$ for $k = 1$, a consequence of dark-count and cross-talk events compensating for photon loss in the waveguide, as per green curve in Fig.~\ref{fig:my_fig_purity}. Furthermore, we find that for a given rate, say 0.1, dark counts are more detrimental to POVM purity than the cross-talk events as evident from black curve ($\delta = 0.1$, $\epsilon =0$) and blue curve ($\delta = 0$, $\epsilon = 0.1$) in Fig.~\ref{fig:my_fig_purity}. In general,  $k-k_{\delta, \epsilon}$ clicks can be mistaken as $k$ clicks, where $k_{\delta, \epsilon}$ are the effective registered clicks due to dark count and cross talk. Thus, it becomes crucial to have a PNR detector with negligible $\delta$ and $\epsilon$ for applications in conditional quantum state preparation and state engineering as well as state characterization with PNR measurements~\cite{Wallentowitz1996, Banaszek1996, Nehra2019, Eaton2019}}
In practice, extremely low dark count rates  have been achieved, which supports our decision to neglect them:
silicon SPADs achieved dark count rates per active area below 1 Hz/$\rm\mu m^2$~\cite{Veerappan2015,Bronzi2015};  InGaAs/InP SPADs tend to have larger dark count rates and 25 $\rm\mu$-diameter devices with a dark count rate of 60 kHz (120 Hz/$\rm\mu m^2$) have been demonstrated~\cite{Jiang2014}. In the particular case of the detection of optical field pulses over much shorter times, it is clear that dark count rates several order of magnitude lower than photon detection rated could be achieved. Moreover, detection of sufficiently short optical pulses will ensure that the dead time due to SPAD quenching can also be ignored.
\section{Conclusion}
 We carried out the theoretical evaluation of the photon-count POVM for a segmented detector such as the one designed in \sec{wg}. Results show that  PNR detection  in the ideal case of no losses and no dark counts requires on the order of $10^{3}$ SPADs to resolve 10 photons, using an efficient gradient coupling scheme. This level of scaling appears to be worthwhile of an integrated optics effort as it would yield a high quality room-temperature PNR detector. While photon losses were taken into account, it is important to note that they did not include the nonideal quantum efficiency of the SPADs, by design of the segmented detector. The reduction of photon losses will therefore only involve passive optical design considerations, a notable difference with terminally coupled tree-splitting detectors in which the quantum efficiency of the SPADs must be unity.  Note also that a tree architecture can still be used to initially split the initial photon number among smaller-sized segmented photodetectors. It is remarkable that reasonable levels of losses (1\% per detector mode), and dark counts and cross-talk noise do not degrade performance as much as having a  limited number of SPADs does, the number of integrated SPADs being then the dominant factor toward high-quality PNR detection.  This means that investing into such a scalable integrated structure, manufacturable with available integrated photonic technology, can yield the benefit of room-temperature high-quality PNR operation. 

 \section{Acknowledgements}
We are grateful to Carlos Gonz\'alez-Arciniegas, Steven van Enk, Jonathan Dowling, Joe Campbell, Seth Bank, Aye L. Win, Rafael Alexander, Ben Godek, Sharon S. Philip, Bargav Jayaraman, and Oshin Jakhete for stimulating discussions. This work was supported by the U.S. Defense Advanced Research Projects Agency and by NSF grants DMR-1839175 and EECS-1842641. 
\bibliographystyle{apsrev4-1}
\bibliography{PNR_ref,Pfister}

% \end{thebibliography}

\end{document}